\documentclass[conference]{IEEEtran}
\IEEEoverridecommandlockouts
\usepackage{cite}
\usepackage{amsmath,amssymb,amsfonts}
\usepackage{algorithmic}
\usepackage{graphicx}
\usepackage{textcomp}
\usepackage{xcolor}
\usepackage{comment}
\usepackage{soul}
\usepackage{multirow}
\usepackage{stfloats}
\usepackage{algorithmic}
\usepackage{subcaption}
\usepackage{booktabs}
\usepackage{soul}
\usepackage{multirow}
\usepackage{stfloats}
\usepackage{algorithmic}
\usepackage{subcaption}
\usepackage{booktabs}
\usepackage{graphicx}
\usepackage{authblk}

\usepackage[ruled,vlined]{algorithm2e}
\usepackage{makecell}   
\usepackage{array}      
\usepackage[margin=1in]{geometry}
\usepackage[ruled,vlined]{algorithm2e}
\usepackage{booktabs}
\def\BibTeX{{\rm B\kern-.05em{\sc i\kern-.025em b}\kern-.08em
    T\kern-.1667em\lower.7ex\hbox{E}\kern-.125emX}}
\usepackage{fancyhdr}
\fancypagestyle{firstpage}
{
    \fancyhead[L]{\footnotesize © 2026 IEEE.  Personal use of this material is permitted.  Permission from IEEE must be obtained for all other uses, in any current or future media, including reprinting/republishing this material for advertising or promotional purposes, creating new collective works, for resale or redistribution to servers or lists, or reuse of any copyrighted component of this work in other works. This paper is accepted at the 12th International Conference on Computing and Artificial Intelligence (ICCAI) 2026.}
    \fancyhead[R]{}
}

\begin{document}
\bstctlcite{IEEEexample:BSTcontrol}

    \title{Sensitivity-Guided Framework for Pruned and Quantized Reservoir Computing Accelerators
}

\author[1]{Atousa Jafari}
\author[2,3]{Mahdi Taheri}
\author[4]{Hassan Ghasemzadeh Mohammadi}
\author[2]{Christian Herglotz}
\author[1]{Marco Platzner}

\affil[1]{Paderborn University, Paderborn, Germany  \{atousa, platzner\}@mail.upb.de}
\affil[2]{Brandenburg Technical University, Cottbus, Germany \{Taheri, Herglotz \}@b-tu.de}
\affil[3]{Tallinn University of Technology, Tallinn, Estonia}
\affil[4]{Reneo Group GmbH, Hamburg, Germany \{info\}@reneogroup.de}

\maketitle
\thispagestyle{firstpage}
\begin{abstract}
This paper presents a compression framework for Reservoir Computing that enables systematic design-space exploration of trade-offs among quantization levels, pruning rates, model accuracy, and hardware efficiency. The proposed approach leverages a sensitivity-based pruning mechanism to identify and remove less critical quantized weights with minimal impact on model accuracy, thereby reducing computational overhead while preserving accuracy. 
We perform an extensive trade-off analysis to validate the effectiveness of the proposed framework and the impact of pruning and quantization on model performance and hardware parameters. For this evaluation, we employ three time-series datasets, including both classification and regression tasks. Experimental results across selected benchmarks demonstrate that our proposed approach is maintaining high accuracy while substantially improving computational and resource efficiency in FPGA-based implementations, with variations observed across different configurations and time series applications. For instance, for the MELBOEN dataset, an accelerator quantized to 4-bit at a 15\% pruning rate reduces resource utilization by 1.2\% and the Power Delay Product (PDP) by 50.8\% compared to an unpruned model, without any noticeable degradation in accuracy.
\end{abstract}


\vspace{-0.4cm}

\section{Introduction}
\label{Introduction}
Reservoir Computing (RC), as a subclass of Recurrent Neural Networks (RNNs), has emerged as a powerful model due to its ability to achieve a performance comparable to other RNNs with significantly lower training complexity and computational overhead.

Recent research has demonstrated how RC can efficiently solve time-series problems and address the drawbacks of traditional RNNs~\cite{tanaka2019recent40}, while achieving comparable accuracy for AI applications, especially for non-linear time series forecasting~\cite{zhang2018spiking47}, pattern classification, and multivariate time series classification~\cite{bloodsample}.

Despite its advantages, RC faces challenges when deployed in real-world scenarios, particularly on resource-constrained edge devices. While RC reduces the training effort compared to traditional RNNs, it often requires numerous neurons and parameters to achieve high accuracy for complex tasks. This increased network size results in significant computational effort and energy requirements during inference, making it difficult to deploy RC models on edge devices such as embedded CPUs, GPUs, or FPGAs. Effective network compression techniques are thus needed to reduce the network size and the computational load without compromising performance~\cite{prune, hawx}. Techniques such as quantization and weight pruning are well-established strategies to address this issue by reducing memory footprint and eliminating redundant connections \cite{huang2023semi}. 

In contrast to prior compressed RC approaches that reduce the number of neurons and connections based on the correlation metric \cite{MI,liu2022broad,huang2023pruning,li2018structure}, our novel sensitivity-guided analysis evaluates how small internal changes influence the model’s performance, capturing the actual functional impact of each neuron on the model's output as well as non-linear and dynamic dependencies. This makes it a more accurate approach and results in less performance degradation. In this work, we develop a sensitivity-guided compression framework to precisely identify critical neurons and prune the reservoir with minimal accuracy loss. We further present a methodology that enables sensitivity-guided design space exploration of both quantization and pruning for FPGA-based RC accelerators. To the best of our knowledge, this is the first framework for an FPGA-based RC model that can achieve not only high throughput and low latency but also be used for design space exploration of trade-offs among quantization levels, pruning rates, model accuracy, and key hardware parameters.

The key contributions of this paper are as follows: 
\begin{itemize}

\item We propose a compression framework that enables the exploration of trade-offs among quantization bit-widths, pruning rates, and hardware implementations, enabling the evaluation of key hardware metrics such as resource utilization, latency, throughput, and power consumption.
\item We employ a novel sensitivity-guided analysis, which identifies and removes the least important neurons on top of the quantized weights to compress the RC network. Additionally, retraining is not required since the weights are quantized. The reduced model capacity can also act as a form of regularization, lowering the risk of overfitting.
\item We implement an automated end-to-end synthesis framework for RC accelerators to map compressed RC models onto FPGAs, leveraging a direct logic implementation, enhancing throughput, latency, and resource efficiency.

\item We conduct a comprehensive analysis for RC FPGA accelerators for both time-series classification and regression tasks.

\end{itemize}

The remainder of this paper is organized as follows: Section~\ref{Background} provides an overview of the RC model as well as prior compression approaches for reservoir computing networks. Section~\ref{Proposed Method} elaborates on the details of the proposed sensitivity-guided framework. Section~\ref{Experimental Results} presents an experimental evaluation, and, finally, Section~\ref{Conclusion} concludes the paper.

\section{Background and Related Work}
\label{Background}
\subsection{Reservoir Computing}
\begin{figure}[htp]
    \centering
\includegraphics[width=0.9\columnwidth, trim={89mm 50mm 80mm 60mm}, clip]{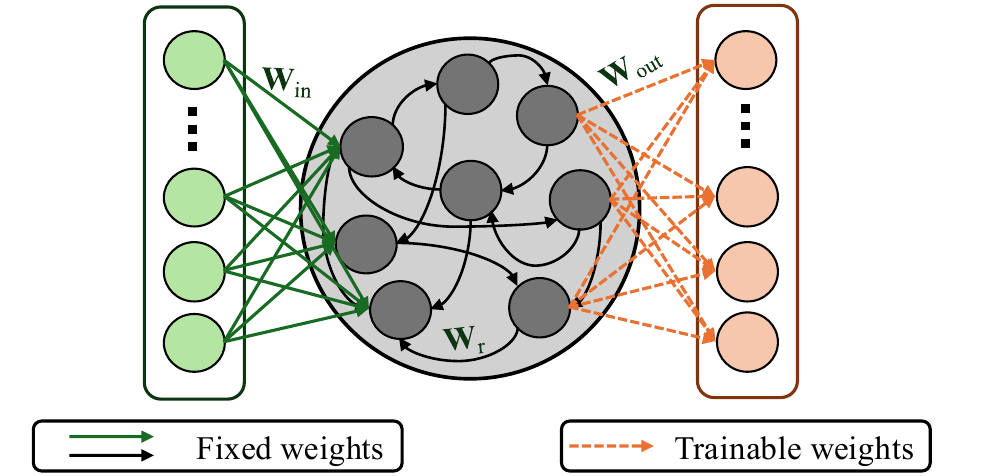}
    \caption{Reservoir  computing architecture consists of three layers: input, reservoir, and output.}
    \label{fig1}
    \vspace{-0.2cm}
\end{figure}

Figure~\ref{fig1} illustrates the typical architecture of reservoir computing (RC). The network consists of three main layers: the input layer, the reservoir layer, and the output layer. The input layer, including neurons, is connected to the reservoir neurons through randomly generated synaptic connections, denoted as
 $\mathbf{W}_{in}$. The reservoir layer contains neurons with sparse interconnections, represented by the matrix $\mathbf{W}_{r}$, which is randomly initialized and remains fixed during training~\cite{tanaka2019recent40}. The output layer is connected to the reservoir via trainable weighted connections, denoted as $\mathbf{W}_{out}$. Notably, only the weights between the reservoir and the output layer require training, which significantly simplifies the training process~\cite{linear,souahlia2020echo_equation}.

The mathematical formulation of RC dynamics follows discrete-time updates. The reservoir state evolution is governed by:

\begin{equation}
    \label{equ:state_update}
    \mathbf{s}(t) = f\left( \mathbf{W}_{in} \mathbf{u}(t) + \mathbf{W}_{r} \mathbf{s}(t-1) \right),
\end{equation}

At any time step $t$, the signals generated by the input layer and reservoir layer are represented by $\mathbf{u}(t)$ and $\mathbf{s}(t)$, respectively. In Equation \ref{equ:state_update}, $\mathbf{W}_{in}$ denotes the input weight matrix, $\mathbf{W}_{r}$ represents the recurrent reservoir weight matrix, and $f$ is typically a saturating non-linear activation function such as $\tanh$. The matrices $\mathbf{W}_{in}$ and $\mathbf{W}_{r}$ are initialized randomly prior to training and remain fixed throughout the network's operation. 
The network output layer ($\mathbf{y}(t)$) is generated through a linear readout:
\begin{equation}
    \label{equ:output}
    \mathbf{y}(t) = \mathbf{W}_{out} \mathbf{s}(t)
\end{equation}
where $\mathbf{W}_{out}$ is the output weight matrix. The training procedure involves determining the optimal $\mathbf{W}_{out}$ that minimizes the difference between the network output and the target signal. 

\subsection{Related Work}
Several prior works have investigated compressing RC networks by selectively removing redundant neurons or connections using correlation-based techniques~\cite{jafari2025crc,MI, liu2022broad, huang2023pruning, li2018structure}. These techniques estimate the neurons' or connections' importance based on statistical dependencies among the reservoir neurons or between reservoir and output neurons. Examples are random pruning, mutual information~\cite{MI}, Spearman correlation~\cite{spearman}, \emph {Principal Component Analysis} (PCA)~\cite{othermethod}, and Lasso~\cite{othermethod}.

However, many of these techniques cannot capture the inherently nonlinear dynamics that define RC systems, making them less suitable for effective pruning. For example, PCA and random pruning are linear transformations and thus cannot capture the nonlinear dynamics of RC. Lasso similarly depends on a linear regression model with $L_1$ regularization, limiting its ability to characterize the nonlinearity of the RC system. Although the mutual information approach can capture nonlinear dependencies, prior works typically use it in an output-unaware manner by evaluating state-to-state correlations rather than the direct influence of a state on the final task's performance~\cite{MI}. Spearman correlation considers nonlinear monotonic dependencies, but it still considers only pairwise dependencies and cannot represent the full complexity of high-dimensional reservoir dynamics~\cite{othermethod}. 
 
A fundamental limitation of correlation-based approaches is the lack of consideration of the complex, nonlinear, and high-dimensional dynamics of RC systems~\cite{othermethod}. Our proposed sensitivity-guided approach, in contrast, directly considers the functional importance of each individual weight by performing simulated bit-flips on quantized weights and measuring the deviation in the output performance $(\mathit{Perf})$ (accuracy/\emph {Root Mean Squared Error} (RMSE)). Our method captures the actual impact of each neuron and corresponding weight on the output performance. This allows for a more precise identification of critical weights compared to methods based on correlation-based approaches. In contrast to prior work, our approach is also quantization-aware, as the sensitivity analysis is performed directly on the quantized weights, leading to less performance degradation during pruning without the retraining requirement, as demonstrated in the experimental results (Section \ref{Results}). Furthermore, we extend our compression framework to generate hardware FPGA accelerators for each compressed model. This allows for studying trade-offs between model performance and hardware-related parameters, which has not been explored in the existing literature.

\section{Proposed Methodology}
\label{Proposed Method}
\begin{figure*}[!htp]
    \centering
    \includegraphics[width=0.8\textwidth, trim={1mm 2mm 1mm 2mm}, clip]{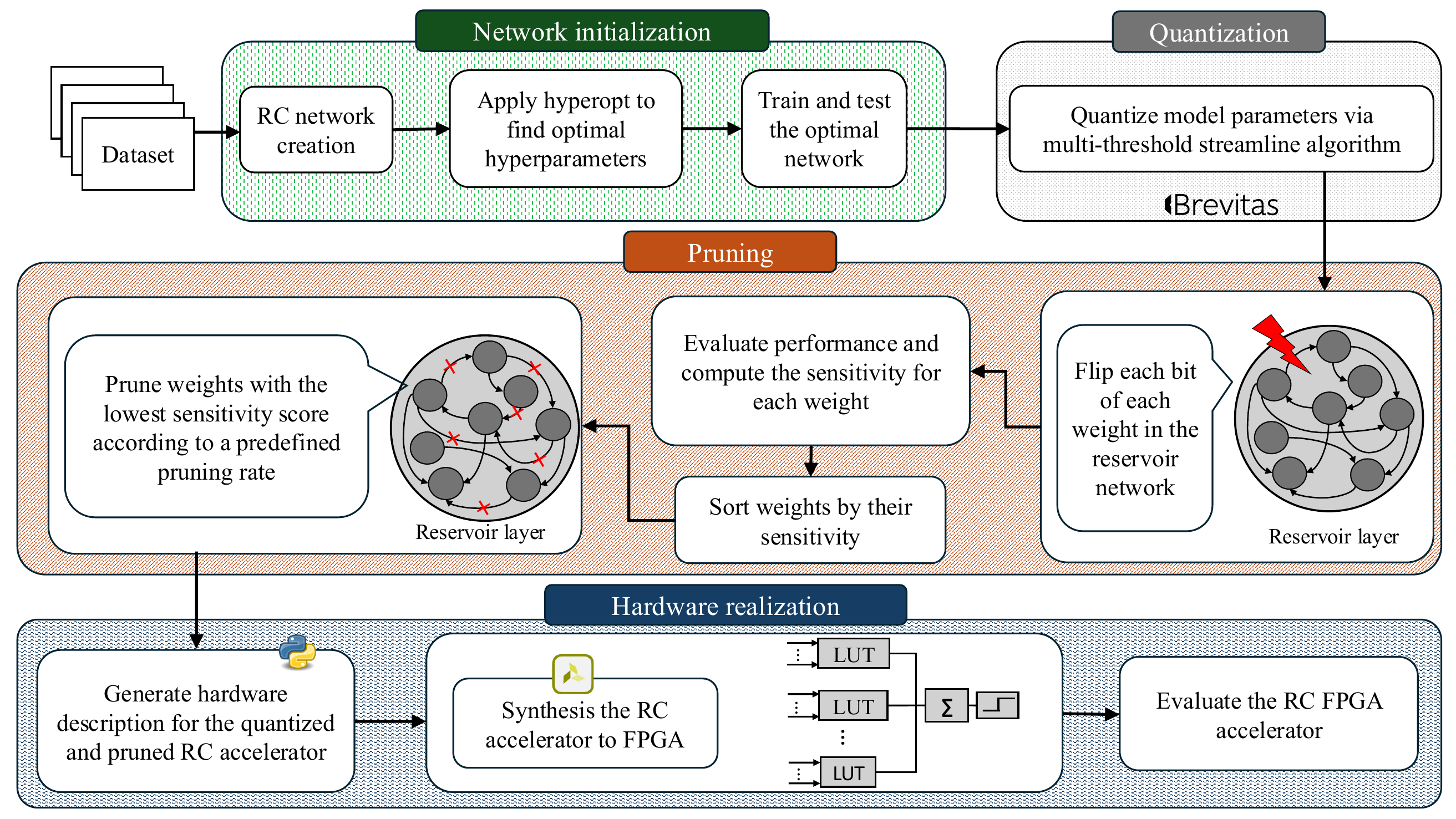}
    \caption{Overview of our RC accelerator synthesis framework, including sensitivity-guided pruning.}
    \label{designflow}
    \vspace{-0.5cm}
\end{figure*}

\subsection{Accelerator Synthesis Flow for RC Networks and Sensitivity-guided Pruning}

Figure~\ref{designflow} illustrates the overall flow of our sensitivity-guided framework for the reservoir computing (RC) model, which consists of four main stages:

The process begins with dataset selection and initialization of configuration parameters. A reservoir model is then created, and optimal hyperparameters are identified, such as spectral radius, leaking rate, sparsity, and the ridge coefficient ($\lambda$) using ReservoirPy’s hyperopt tool~\cite{trouvain2020reservoirpy}. Once the best-performing configuration is identified based on the user’s performance requirements, the model is subsequently passed to the next stage.

In the quantization stage, linear quantization is employed as follows: 
\begin{equation}
   x_{int} = scale \times (x-b)
    \label{equ:quant}
\end{equation}

In Equation~\ref{equ:quant}, $x$ is the original floating-point number, $scale$ and $b$ are the scaling factor and chosen bias in floating-point format, and $x_{int}$ is the $q$-bit quantized integer for $x$. For a hardware-friendly deployment, the streamline algorithm~\cite{Stream} is employed, which absorbs floating-point operations (e.g., scale, bias) extracted during the linear quantization process into the activation function~\cite{Stream,jafari2025ultra}. In particular, the quantized activation function layer (in our case, $HardTanh$) is converted into successive multi-threshold integer steps, which relate to the bit-width quantization. Each input value is compared with the threshold and mapped to the nearest index. We denote the model performance (accuracy, RMSE) after $q$-bit quantization as 
$\mathit{Perf}^{base}(q)$.

In the pruning stage, we identify and remove less critical reservoir weights based on their sensitivity. To this aim, we induce bit-flips (0→1 and 1→0) for each bit position $b$ of each quantized weight $w \in W_r^{(q)}$. Then, we measure the impact of each single bit-flip on model accuracy and determine the resulting performance as $\mathit{Perf}^{b,w}(q)$. The bit-flips can be seen as fault injections~\cite{rakin2019bit} and allow us to investigate the effect on performance when we deviate weights. Finally, the sensitivity score of a weight $w$ is computed as the average deviation across all bit positions (high bits vs. low bits) of the weight: 
\begin{equation}
\mathit{Sensitivity}(w) = \frac{1}{q} \sum_{b=1}^{q} | \mathit{Perf}^{base}(q) - \mathit{Perf}^{b,w}(q) |
\label{equ:score}
\end{equation}

Weights with low sensitivity scores are less critical to the model's output and are therefore ideal candidates for pruning.
Once we have computed all sensitivity scores, the quantized weights are ranked in ascending order. The pruning stage of our flow proceeds by removing the first $p\%$ of the weights, i.e., the $p\%$ weights with the lowest sensitivity scores, where $p$ is denoted as the pruning rate. 


The hardware realization stage first converts the RC model in configuration $s(q,p)$ to a Register-Transfer Level (RTL) description by means of a Python script. The generated RTL is then synthesized to an FPGA for evaluating the design in terms of performance (accuracy/RMSE), resource utilization, latency and throughput, and power consumption. Our accelerator designs are based on the direct logic implementation approach. In this custom-tailored method, all layers of the RC network are mapped into Lookup Tables (LUTs) structures. Since the weights of the RC model are known, they can be hardwired into the LUTs, avoiding costly memory fetches and stores. Furthermore, multiplication operations are converted into equivalent shift/add operations. Such mapping ensures low-latency and high-throughput accelerators~\cite{logicnets,directlogic}.

\subsection{Design Space Exploration}

The flow in Figure~\ref{designflow} can be easily modified and extended to create hardware accelerators for RC networks with given bounds on hardware resources or bounds on acceptable degradations in accuracy/RMSE, respectively. In this work, we use the flow to conduct a systematic design space exploration (DSE) over different compressions, i.e., quantization and pruning.
Algorithm~\ref{alg:reliability_fault_pruning} outlines the DSE procedure. The algorithm extends the quantization and pruning stages in Figure~\ref{designflow} and is provided with a set of quantizations ($q\in Q$) and pruning rates ($p \in P$) that should be applied. The result is a set of accelerator configurations ($s(q,p) \in S$) that then can be passed on to the hardware realization stage. 

\begin{algorithm}[!htp]
\caption{Design space exploration over quantization and pruning stages.}
\label{alg:reliability_fault_pruning}
\LinesNumbered 
\DontPrintSemicolon
\KwIn{Set of quantization bit-widths $Q$, set of pruning rates $P$}
\KwOut{Set of accelerator configurations $S$ }

$S \gets \emptyset $ \;
\ForEach{ $q \in Q$}{
    Quantize $W_r$ to obtain $W_r^{(q)}$\;
    Evaluate $\mathit{Perf}^{base}(q)$\;
    \ForEach{weight $w \in W_r^{(q)}$}{
        \For{$b = 1$ \KwTo $q$}{
            Flip bit $b$ of $w$ and evaluate  $\mathit{Perf}^{b, w}$\;
     }
 \[
\begin{split}
\mathit{Sensitivity}(w) = \\
\frac{1}{q} \sum_{b=1}^{q} 
\left| \mathit{Perf}^{base}(q) - {} \right.
\left. \mathit{Perf}^{b,w}(q) \right|
\end{split}
\]
            

    }
    Sort all weights in ascending order of $\mathit{Sensitivity}(w)$\;

    \ForEach {$p \in P$}{
        Prune the lowest $p\%$ of weights\;
        Measure $\mathit{Perf}^{(p,q)}$\;
        Create accelerator configuration $s(q,p)$ \;
        $S  \gets S \cup s(q,p)$ \;
    }
 }

\Return{$S$}\;

\end{algorithm}
The algorithm iterates over all quantization levels $q$ (line 2). After the design is quantized and the baseline performance determined (lines 3-4), iteration over all weights of the model and over all bits of each weight is performed (lines 5-6). The selected bits are flipped, and the corresponding performance is determined (line 7). For each weight, the sensitivity score is computed (line 8). After sorting the weights in ascending order of sensitivity (line 9), we iterate over all specified pruning rates (line 10), prune the network accordingly (line 11), and measure the resulting performance $\mathit{Perf}^{(p,q)}$ of the quantized and pruned accelerator. Finally, the accelerator is added to the set of configurations for subsequent hardware realization.


This DSE enables a thorough investigation of compression trade-offs, in particular how different quantization levels and pruning rates affect model performance, resource utilization, latency, throughput, and energy efficiency. Moreover, the DSE allows for comparisons of the sensitivity-guided pruning approach with other known pruning techniques.  

\color{black}
\begin{figure*}[!htp]
    \captionsetup[subfigure]{labelformat=empty}
    \centering

    \begin{subfigure}{0.3\textwidth}
        \centering
        \includegraphics[width=\linewidth]{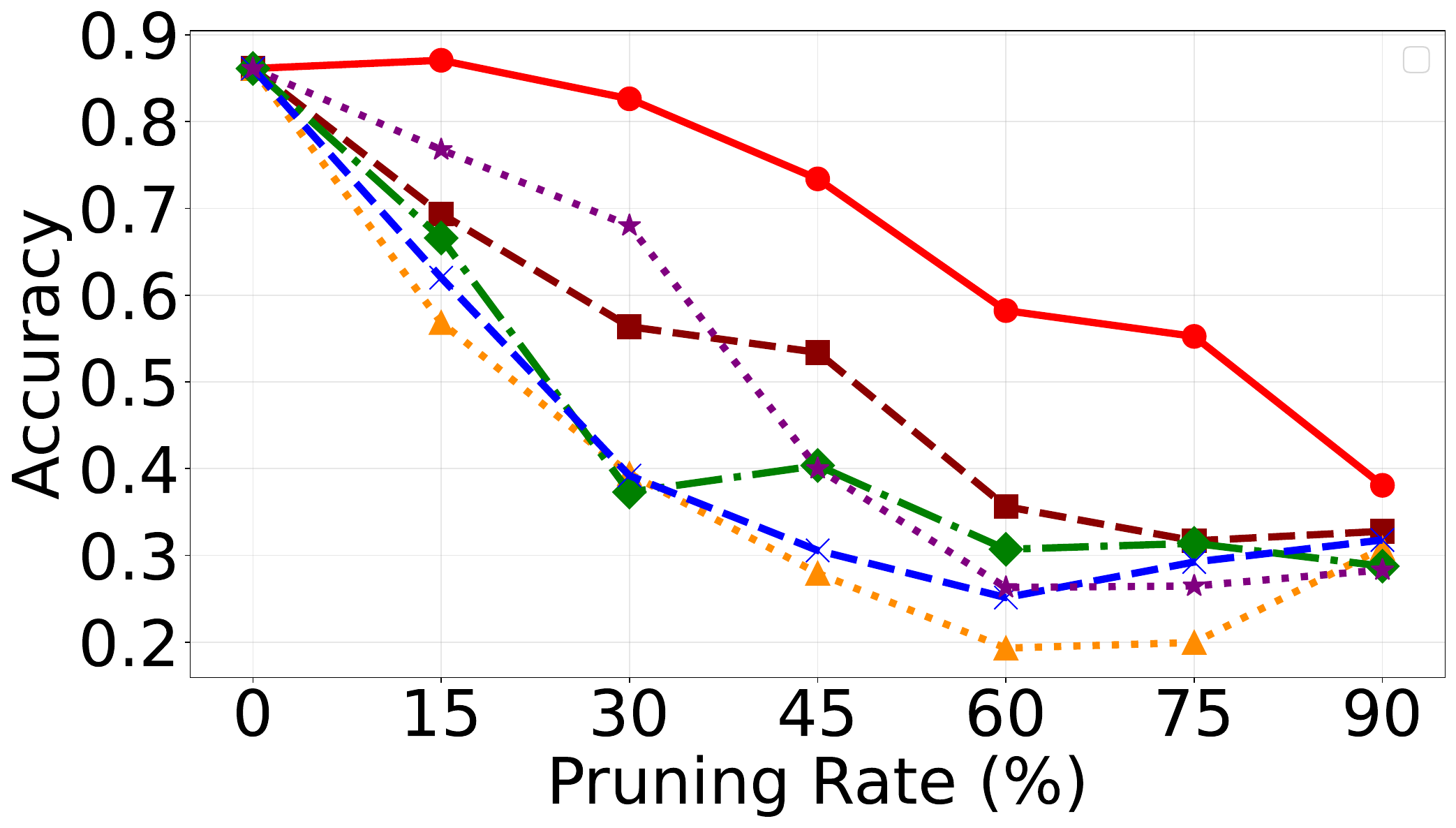}
        \caption{4-bit MELBORN}
    \end{subfigure}
    \hfill
    \begin{subfigure}{0.3\textwidth}
        \centering
        \includegraphics[width=\linewidth]{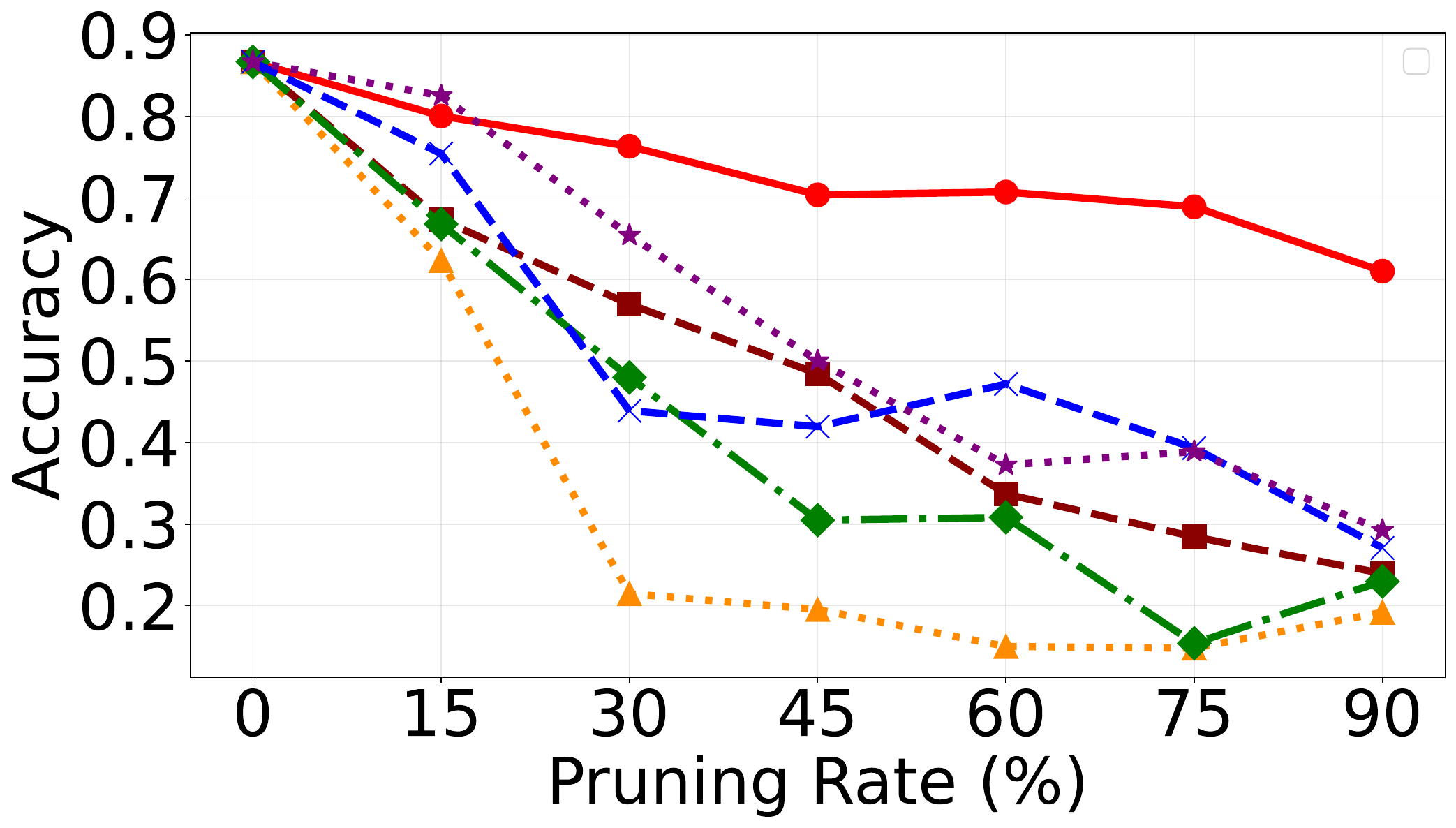}
        \caption{6-bit MELBORN}
    \end{subfigure}
    \hfill
    \begin{subfigure}{0.3\textwidth}
        \centering
        \includegraphics[width=\linewidth]{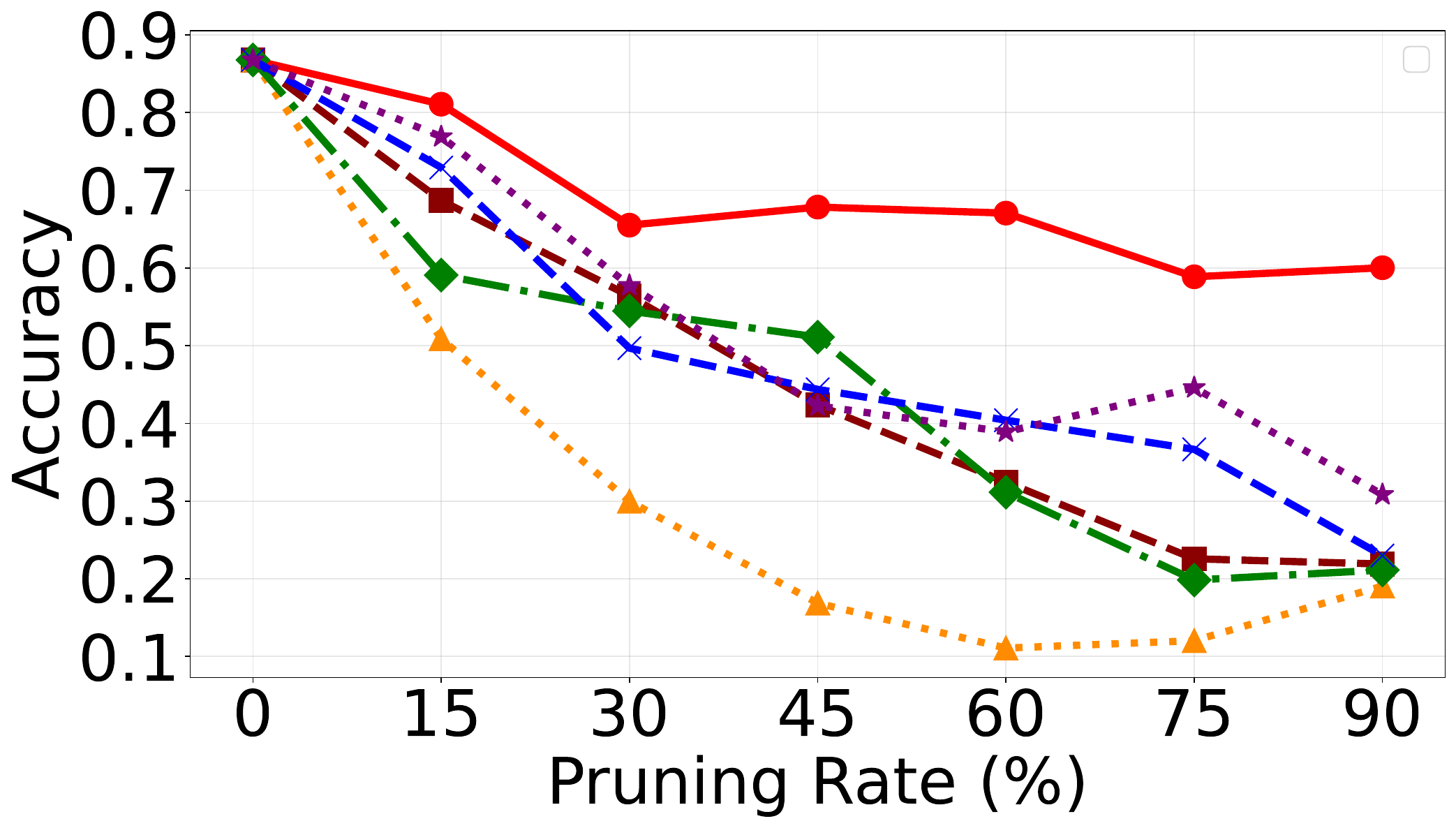}
        \caption{8-bit MELBORN}
    \end{subfigure}

    \begin{subfigure}{0.3\textwidth}
        \centering
        \includegraphics[width=\linewidth]{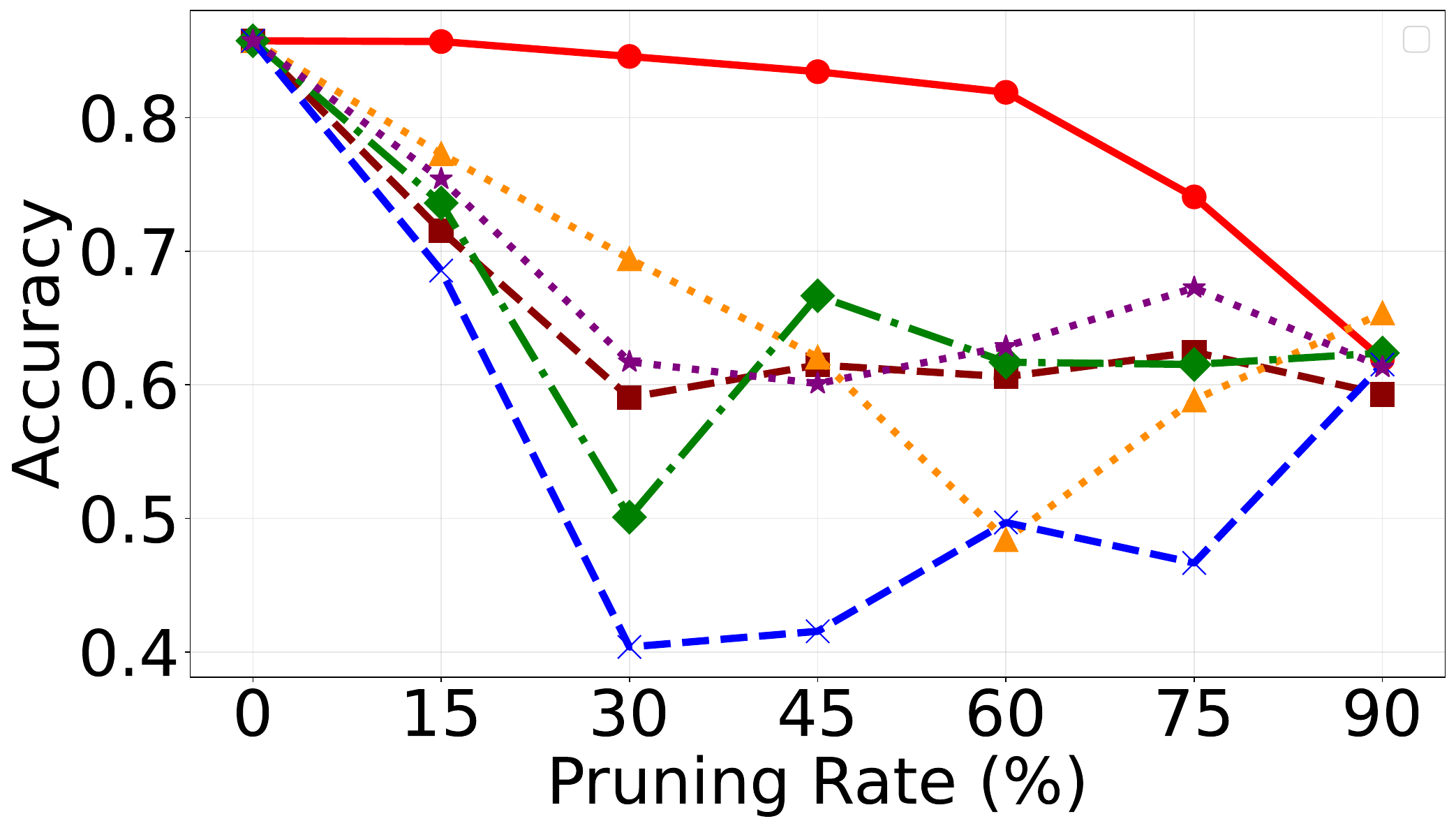}
        \caption{4-bit PEN}
    \end{subfigure}
    \hfill
    \begin{subfigure}{0.3\textwidth}
        \centering
        \includegraphics[width=\linewidth]{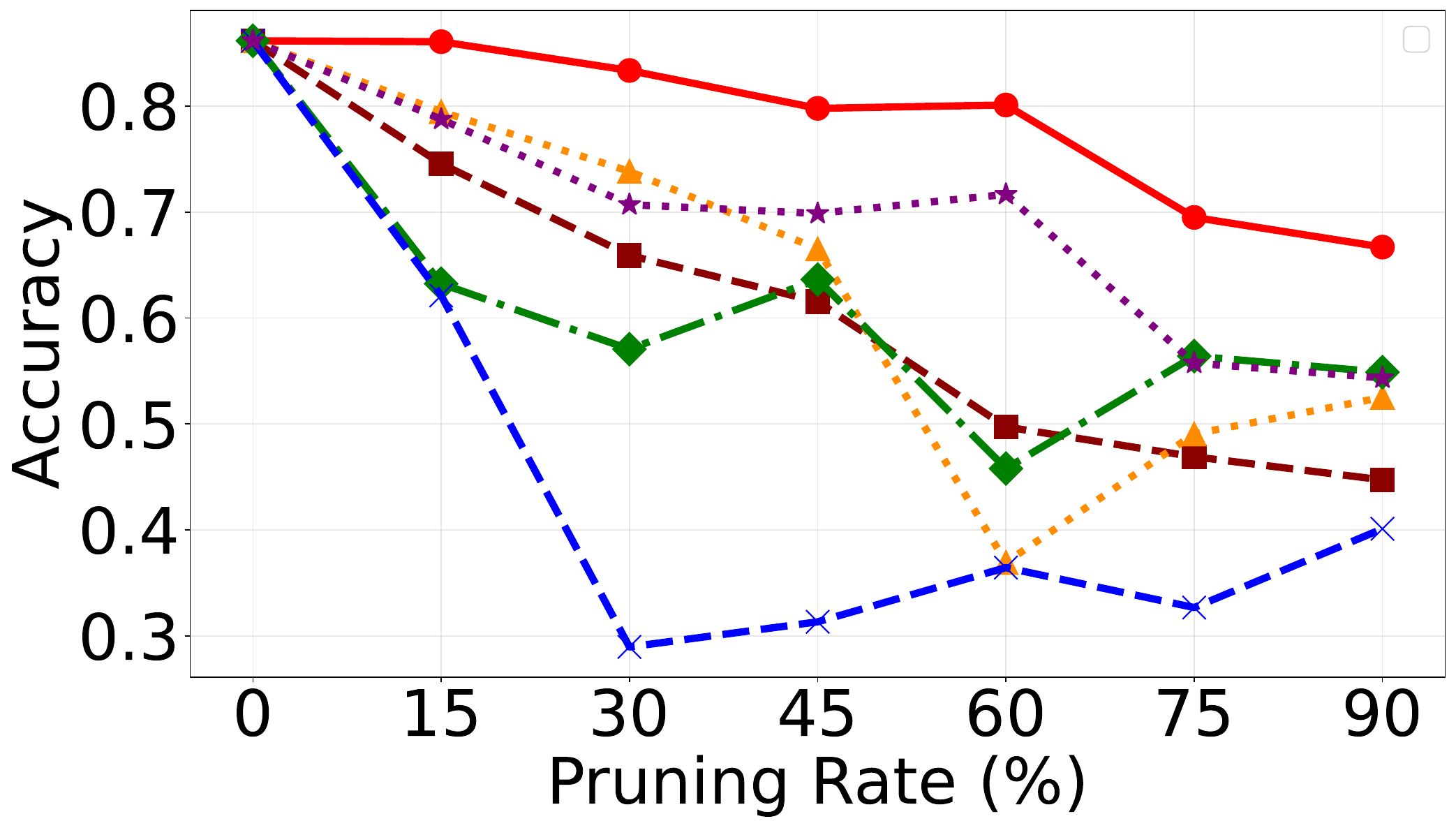}
        \caption{6-bit PEN}
    \end{subfigure}
    \hfill
    \begin{subfigure}{0.3\textwidth}
        \centering
        \includegraphics[width=\linewidth]{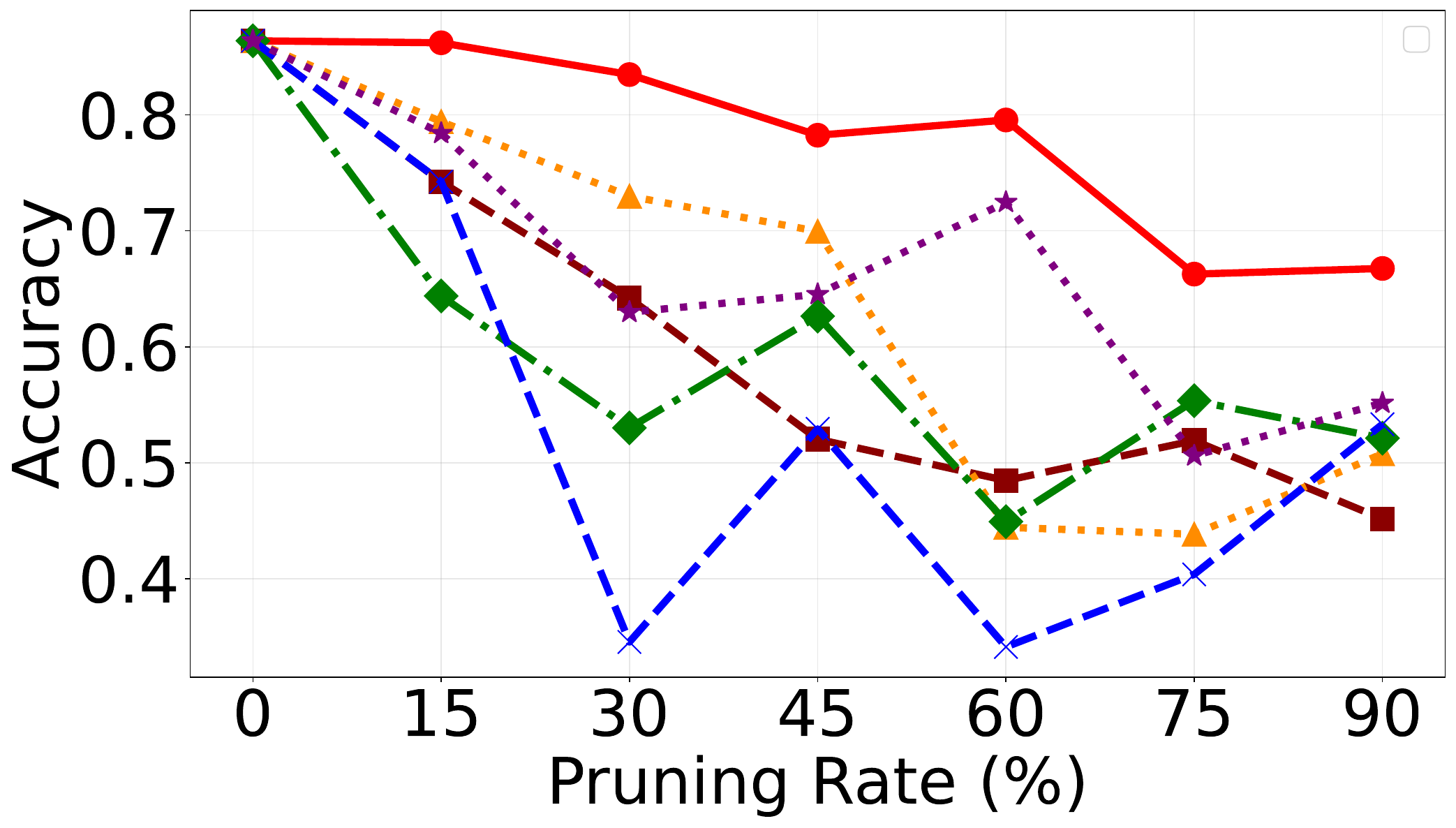}
        \caption{8-bit PEN}
    \end{subfigure}

    \begin{subfigure}{0.3\textwidth}
        \centering
        \includegraphics[width=\linewidth]{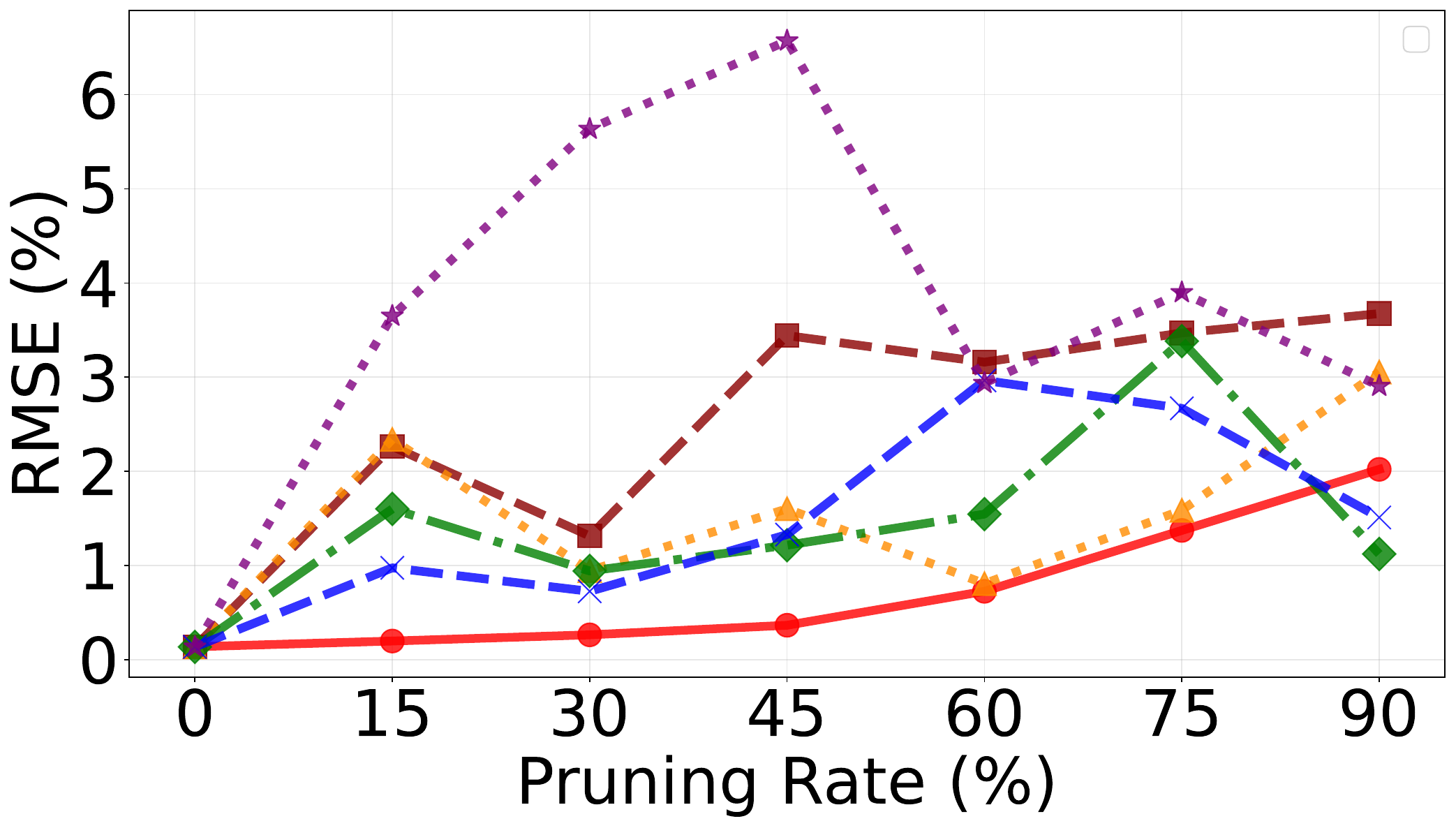}
        \caption{4-bit HENON}
    \end{subfigure}
    \hfill
    \begin{subfigure}{0.3\textwidth}
        \centering
        \includegraphics[width=\linewidth]{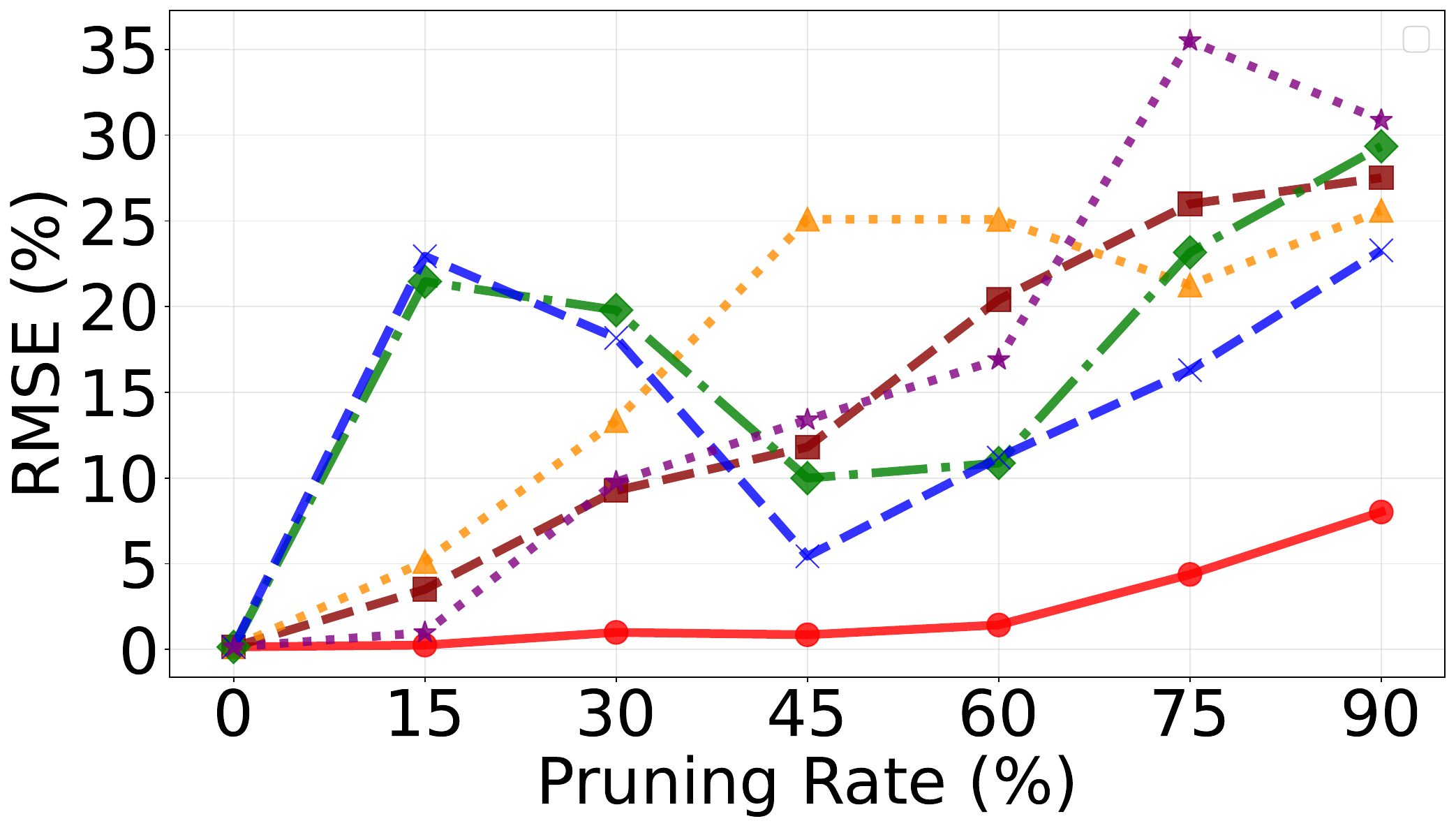}
        \caption{6-bit HENON}
    \end{subfigure}
    \hfill
    \begin{subfigure}{0.3\textwidth}
        \centering
        \includegraphics[width=\linewidth]{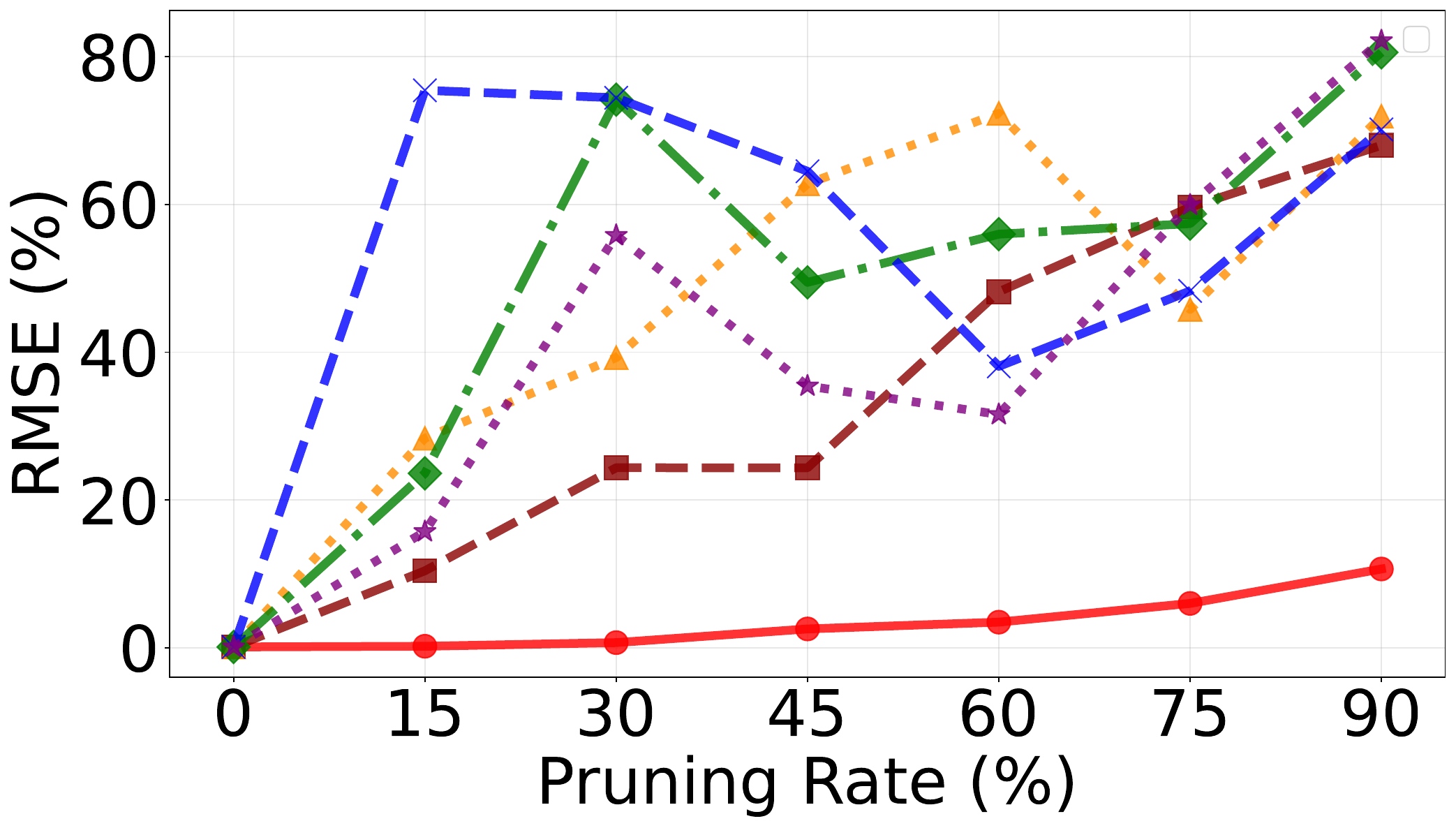}
        \caption{8-bit HENON}
    \end{subfigure}

    \vspace{0.3cm}
    \begin{minipage}{0.95\textwidth}
        \centering
        \includegraphics[width=0.8\textwidth]{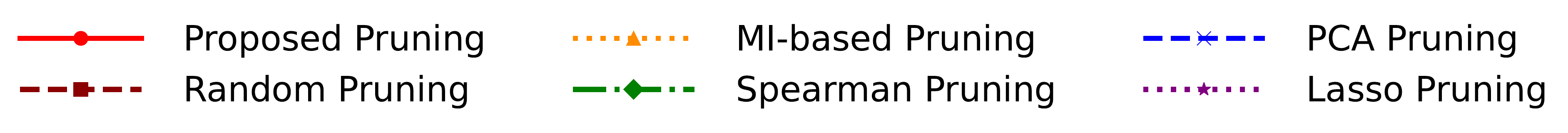}
    \end{minipage}

    \caption{Performance evaluation for various quantization levels and pruning rates across selected time-series datasets.}
    \label{fig:performance evaluation}
   \vspace{-0.5cm}
\end{figure*}

\section{Experimental Results}
\label{Experimental Results}
\subsection{Reservoir Computing Benchmarks}
We use several time-series datasets for experimentation: MELBORN and PEN as classification tasks and HENON as a regression task. 

\begin{table}[!htp]
\centering
\footnotesize
\caption{Parameters for time-series benchmarks.}
\label{tab:datasets}
\small
\begin{tabular}{|l|c|c|c|}
\hline
                   & \multicolumn{3}{c|}{\textbf{Benchmark}} \\ \hline
\textbf{Parameter} & \textbf{MELBORN}  & \textbf{PEN} & \textbf{HENON}  \\ \hline
$N$                & 50    & 50    & 50        \\ \hline
$S_{length}$       & 24    & 8     & 5000      \\ \hline
$\#$classes        & 1     & 10    & (regression) \\ \hline
$T_{train}$        & 1194  & 7494  & 4000      \\ \hline
$T_{test}$         & 2439  & 3498  & 1000      \\ \hline
$sr, lr, \lambda$  & 0.9, 1, 1e-11 & 0.6, 1, 1e-5 & 0.9, 1, 1e-8 \\ \hline
$ncrl$             & 250   & 250   & 250       \\ \hline
$\mathit{Perf}$    & 87.67\% & 86.34\% & 0.27\%$^{\dagger}$ \\ \hline
\end{tabular}

\vspace{1mm}
\footnotesize
\raggedright
$^{\dagger}$Performance for the HENON benchmark is reported as RMSE.
\vspace{-0.4CM}
\end{table}

For each task, we employ the first stage of our flow in Figure~\ref{designflow} and leverage ReservoirPy’s hyperopt tools to conduct a random search between 1000 different sets of hyperparameters. Table~\ref{tab:datasets} presents the resulting hyperparameters for all three benchmarks.

The table lists the number of reservoir neurons ($N$), the input sequence length $S_{length}$, the number of classes ($\#$classes), the sizes of the training and test data ($T_{train}$, $T_{test}$), the spectral radius ($sr$), the leaking rate ($lr$), the ridge coefficient ($\lambda$), the numbers of weight connectivity in the reservoir layer ($ncrl$), and the original performance (unquantized/unpruned models).


\vspace{0.1cm}
\begin{table*}[!htbp]
\centering
\caption{Evaluation of hardware utilization, latency, throughput, and PDP for quantized and (sensitivity-guided) pruned RC FPGA accelerators using the MELBORN dataset.}
\label{tab:MELBORN}
\begin{tabular}{ccccccccc}
\hline
\textbf{Quantization} & \textbf{Pruning} & \multicolumn{2}{c}{\textbf{Resource Utilization}} & \textbf{Latency (ns)} & \textbf{Throughput} & \textbf{PDP } & \textbf{\textbf{Resource Saving }} & \textbf{\textbf{PDP Saving}} \\ 
\cline{3-4} \textbf{$q$  Bits } & \textbf{rate $p$ } & \textbf{LUTs} & \textbf{FFs} & (ns) & (Msps) & (nWs) & (\%) & (\%) \\ 
\hline
 & Unpruned & 29400 & 558 & 16.220 & 61.67 & 9.408 & \textbf{-} & \textbf{-}\\
 & 15\% & 29027 & 543 & 8.343 & 119.86 & 4.618 & \textbf{1.26} & \textbf{50.88}\\ 
 4 & 45\% & 28728 & 516 & 7.040 & 142.05 & 3.772 & \textbf{2.27} & \textbf{59.91}\\ 
 & 75\% & 28227 & 516 & 4.380 & 228.31 & 2.342 & \textbf{3.99} & \textbf{75.09}\\ 
 & 90\% & 28170 & 508 & 4.278 & 233.80 & 2.228 & \textbf{4.17} & \textbf{76.31}\\ 
\hline
 & Unpruned & 42893 & 339 & 9.962 & 100.38 & 6.773 & \textbf{-} & \textbf{-}\\ 
  & 15\% & 42067 & 335 & 8.550 & 116.95 & 5.393 & \textbf{1.93} & \textbf{20.39}\\ 
 6 & 45\% & 40470 & 325 & 8.732 & 114.48 & 5.695 & \textbf{5.61} & \textbf{15.91}\\ 
 & 75\% & 38920 & 320 & 8.795 & 113.73 & 5.409 & \textbf{9.23} & \textbf{20.13}\\ 
 & 90\% & 38453 & 317 & 7.232 & 138.22 & 4.389 & \textbf{10.29} & \textbf{35.20}\\ 
\hline
 & Unpruned & 63208 & 400 & 10.795 & 92.62 & 8.636 & \textbf{-} & \textbf{-}\\ 
  & 15\% & 61578 & 400 & 10.056 & 94.69 & 7.480 & \textbf{2.56} & \textbf{13.39}\\ 
 8 & 45\% & 58493 & 400 & 9.564 & 104.56 & 7.153 & \textbf{7.41} & \textbf{17.17}\\ 
 & 75\% & 55173 & 400 & 8.555 & 116.91 & 6.041 & \textbf{12.64} & \textbf{30.04}\\ 
 & 90\% & 54386 & 400 & 7.742 & 129.12 & 5.358 & \textbf{13.89} & \textbf{37.93}\\ 
\hline
\end{tabular}
\end{table*}

\label{Results}

\subsection{Performance of Sensitivity-guided Pruning}
Using the DSE procedure of Algorithm~\ref{alg:reliability_fault_pruning} with quantizations $Q=\{4,6,8\}$ bits and pruning rates $P=\{15, 30, 45, 60, 75, 90\}$ percent, we have generated the results illustrated in Figure~\ref{fig:performance evaluation}. The subfigures show the model performance, i.e., accuracy for the classification tasks MELBORN and PEN (higher is better) and RMSE for the regression task HENON (lower is better), over the pruning rate. In each subfigure, the baseline (unpruned) performance is represented by a dotted line. 

For a comparative performance analysis of the sensitivity-guided pruning against other known pruning techniques from literature, we have also conducted the experiments with random pruning, MI-based~\cite{MI}, Spearman rank correlation, PCA, and Lasso~\cite{othermethod} across all quantization levels. For a fair comparison, we evaluate all pruning techniques at the same pruning rates.

Figure~\ref{fig:performance evaluation} clearly shows that our proposed sensitivity-guided pruning surpasses conventional pruning approaches. The only exceptions are for PEN and HENON quantized to 4-bit at a 90\% pruning rate. But also here, sensitivity-guided pruning is among the best techniques. Additionally, for classification tasks, the conventional pruning techniques exhibit a more significant accuracy drop with more aggressive pruning. For instance, in the MELBORN dataset, accuracy remains above 0.6 even at 60–75\% pruning with our proposed pruning method, while all other techniques drop below 0.4. Similarly, for the regression task (HENON), our proposed method not only achieves the lowest RMSE but also demonstrates more gradual performance degradation across pruning rates. Overall, the results show that compared to known techniques from literature, our proposed sensitivity-based pruning technique leads to i) a smaller performance drop and ii) exhibits a less rapidly declining performance across a wide range of pruning rates and quantization levels.

\subsection{RC Accelerator Synthesis and Evaluation}
  
\begin{table*}[!htbp]
\centering
\caption{Evaluation of hardware utilization, latency, throughput, and PDP for quantized and (sensitivity-guided) pruned RC FPGA accelerators using the HENON dataset.}
\label{tab:HENON}
\begin{tabular}{ccccccccc}
\hline
\textbf{Quantization} & \textbf{Pruning} & \multicolumn{2}{c}{\textbf{Resource Utilization}} & \textbf{Latency } & \textbf{Throughput } & \textbf{PDP (nWs)} & \textbf{Resource Saving} & \textbf{PDP Saving} \\ 
\cline{3-4} \textbf{$q$ Bits} & \textbf{rate $p$ } & \textbf{LUTs} & \textbf{FFs} & (ns) & (Msps) & (nWs) & (\%) & (\%) \\ 
\hline
 & Unpruned & 3448 & 196 & 5.581 & 179.2 & 0.341 & \textbf{-} & \textbf{-}\\
 & 15\% & 3112 & 190 & 5.378 & 185.9 & 0.301 & \textbf{9.38} & \textbf{11.73}\\ 
 4 & 45\% & 2426 & 176 & 4.383 & 228.2 & 0.202 & \textbf{26.36} & \textbf{40.76}\\ 
 & 75\% & 1971 & 134 & 2.984 & 335.2 & 0.116 & \textbf{40.42} & \textbf{65.98}\\ 
 & 90\% & 1869 & 76 & 2.616 & 382.3 & 0.094 & \textbf{51.63} & \textbf{72.44}\\ 
\hline
 & Unpruned & 7102 & 300 & 7.286 & 137.2 & 0.707 & \textbf{-} & \textbf{-}\\ 
 & 15\% & 6314 & 293 & 7.012 & 142.6 & 0.624 & \textbf{10.32} & \textbf{11.74}\\ 
 6 & 45\% & 4501 & 268 & 6.537 & 153.0 & 0.445 & \textbf{34.51} & \textbf{37.06}\\ 
 & 75\% & 2869 & 197 & 5.201 & 192.2 & 0.265 & \textbf{59.18} & \textbf{62.52}\\ 
 & 90\% & 2175 & 114 & 2.834 & 352.9 & 0.125 & \textbf{73.20} & \textbf{82.32}\\ 
\hline
 & Unpruned & 11469 & 400 & 8.251 & 121.2 & 1.016 & \textbf{-} & \textbf{-}\\ 
 & 15\% & 9854 & 400 & 7.901 & 126.6 & 0.878 & \textbf{12.64} & \textbf{13.56}\\ 
 8 & 45\% & 7005 & 376 & 7.185 & 139.2 & 0.604 & \textbf{36.83} & \textbf{40.55}\\ 
 & 75\% & 3909 & 295 & 6.420 & 155.8 & 0.373 & \textbf{64.64} & \textbf{63.27}\\ 
 & 90\% & 2424 & 192 & 3.831 & 261.1 & 0.176 & \textbf{81.36} & \textbf{82.67}\\ 
\hline
\end{tabular}
\vspace{-0.5cm}
\end{table*}

We generate RC accelerators with 4-, 6-, and 8-bit quantization for both the unpruned model and the models pruned using the sensitivity-based technique at the pruning rates shown in Figure~\ref{fig:performance evaluation}. The accelerators have been synthesized with AMD/Xilinx Vivado 2022.2 to the Virtex Ultra-Scale xcvu19p-fsvb3824-1-e device. Post-place-and-route power estimates have been generated by Vivado using Switching Activity Interchange Format (SAIF) files obtained from post-synthesis simulation of each benchmark. These SAIF files have been subsequently injected into Vivado for detailed power estimation.  


Tables \ref{tab:MELBORN} and \ref{tab:HENON} provide a detailed analysis of how quantization and the proposed pruning jointly affect the hardware performance of the RC accelerator deployed on FPGA. The evaluation metrics include resource utilization (LUTs and FFs), latency, throughput, and the Power–Delay Product (PDP). Since our accelerators adopt the direct logic implementation approach, utilizing only LUTs and FFs without any Block RAM usage, they eliminate memory-access bottlenecks, which results in ultra-low latency and extremely high throughput designs. The resource (LUTs+FFs) and PDP savings are calculated with respect to the quantized but unpruned RC baseline.

Tables~\ref{tab:MELBORN} and~\ref{tab:HENON} further demonstrate the trade-off between quantization bit-width, pruning rate, and hardware efficiency. Generally, lower quantization bit-widths and higher pruning rates reduce hardware cost and, even more pronounced, energy requirements. This trade-off holds across both datasets. For instance, for the MELBORN classification task at 4-bit precision and 15\% pruning rate, we achieve a 50.88\% reduction in PDP and 1.26\% resource saving while maintaining the accuracy (cmp.~Figure~\ref{fig:performance evaluation}) compared to the unpruned model.

\begin{figure*}[!htp]
    \centering
    \begin{minipage}{0.45\textwidth}
        \centering
        \includegraphics[width=\textwidth]{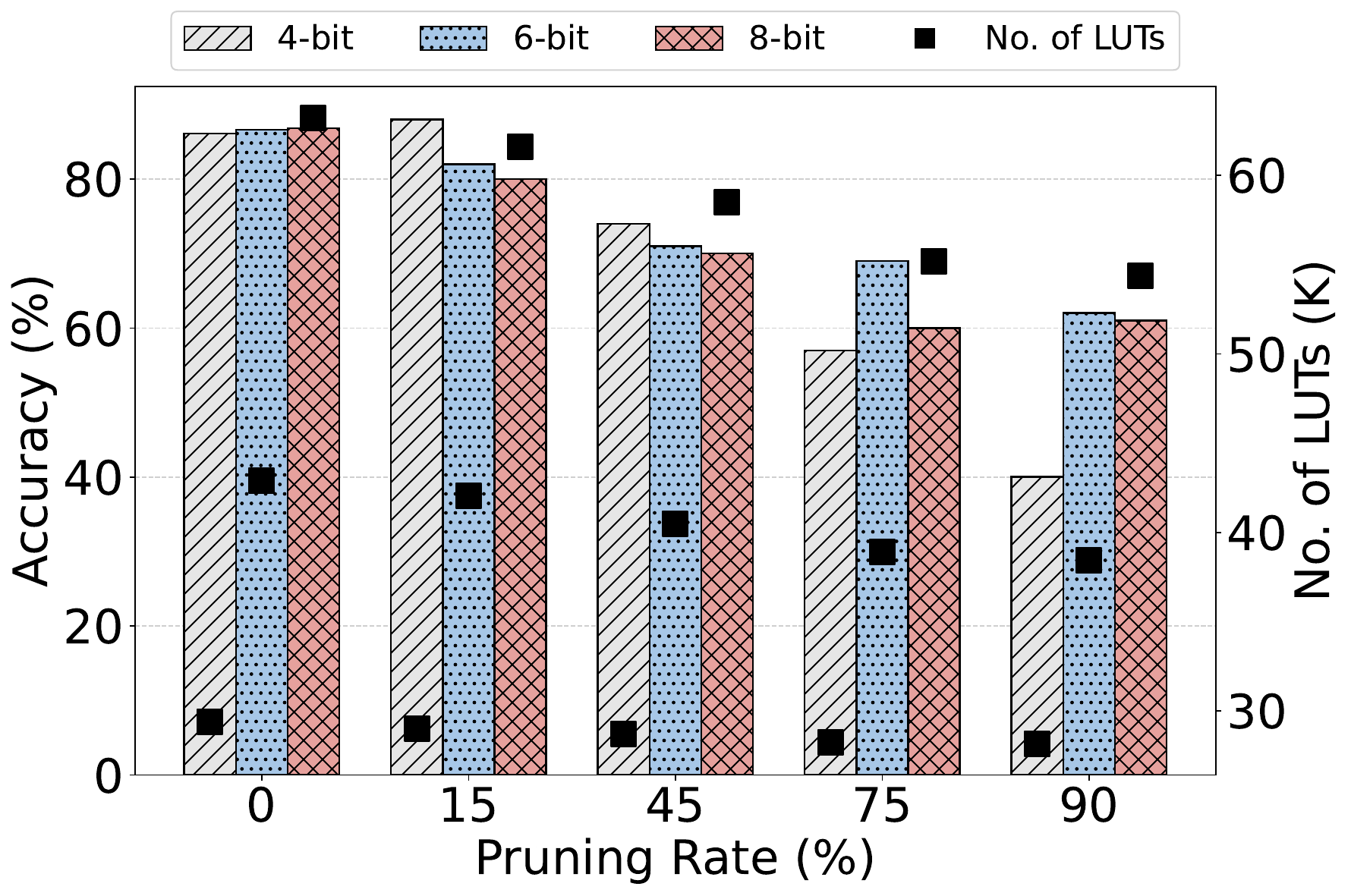}
        \subcaption{MELBORN}
        \label{fig:DSE_Melborn}
    \end{minipage}
    \hfill
    \begin{minipage}{0.45\textwidth}
        \centering
        \includegraphics[width=\textwidth]{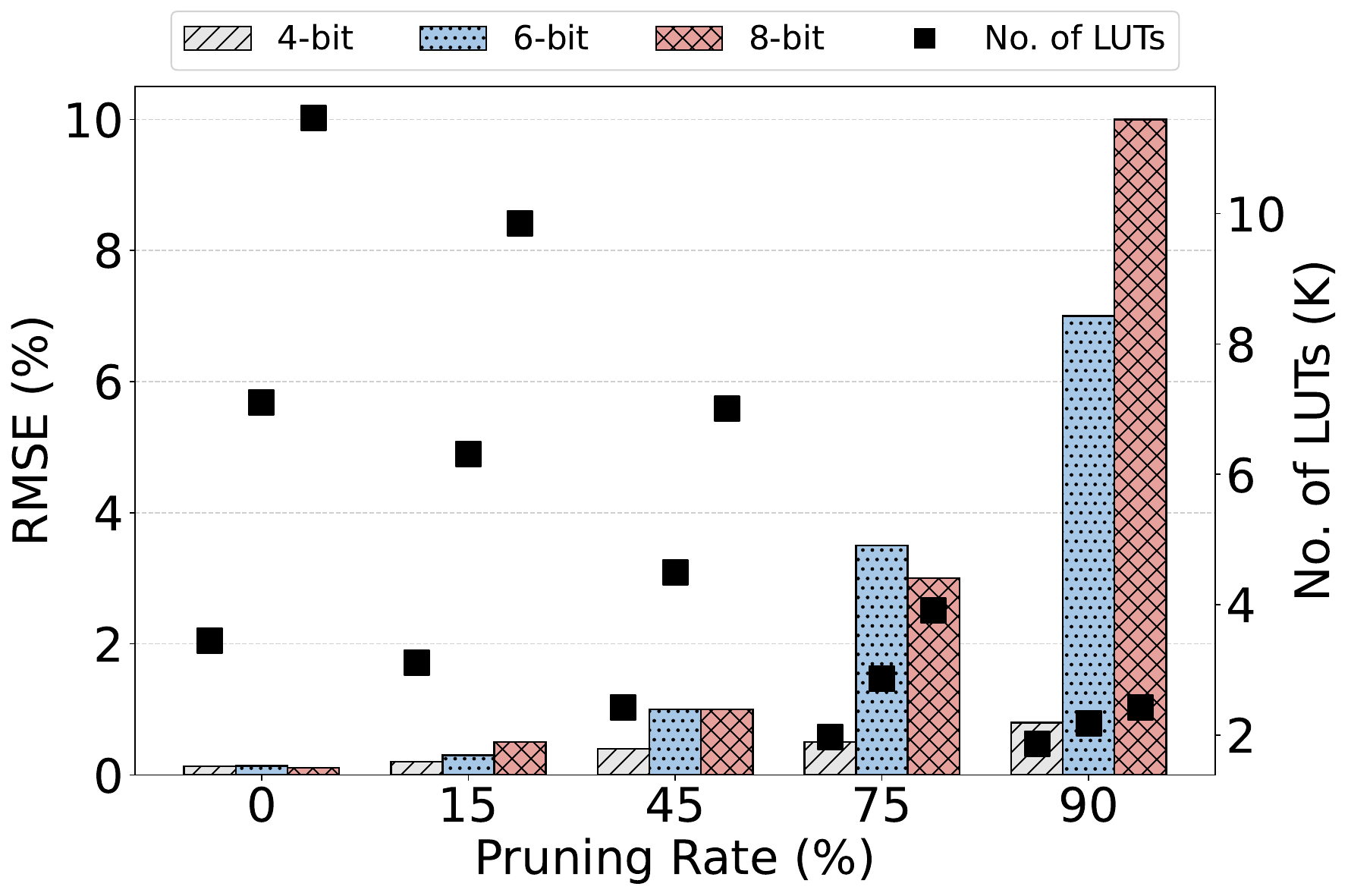}
        \subcaption{HENON}
        \label{fig:DSE_Henon}
    \end{minipage}
    \caption{Trade-off analysis between performance and resource consumption for quantized and pruned RC accelerators.}
    \label{fig:DSE}
    \vspace{-0.6cm}
\end{figure*}

Figure~\ref{fig:DSE} combines accuracy/RMSE data from Figure~\ref{fig:performance evaluation} with results from hardware generation for different quantization levels and pruning rates. Notably, while increasing quantization levels and pruning rates saves on hardware resources, the dependency of the accuracy/RMSE on quantization levels and pruning rates is more involved. As shown in the figure, going down from 8 over 6 to 4-bit quantization at a 15\% pruning rate can even improve accuracy or reduce RMSE, respectively. This highlights the importance of conducting such detailed design space exploration to find the best accelerator configuration for a given dataset.


\section{Conclusion}
\label{Conclusion}
In this paper, we have proposed a framework that leverages sensitivity-based pruning and quantization to compress RC models for efficient hardware implementation. The proposed framework is capable of applying various bit-width quantizations and pruning rates and generating the corresponding hardware implementation for each configuration. This enables the exploration and selection of the optimal trade-offs between accuracy, resource utilization, latency/throughput, and energy efficiency for a given dataset. Our proposed framework will also be made available as open-source. Further work includes the mapping of more complex and larger RC models, such as deep Echo State Networks (ESN). 

\color{black}
\section*{Acknowledgment}
\scriptsize
This research is funded in part by the German Federal Ministry for the Environment, Nature Conservation, Nuclear Safety, and Consumer Protection under grant no. 67KI32004A, by the Estonian Research Council grant PUT PRG1467”CRASHLESS“, EU Grant Project 101160182 “TAICHIP“, by the Deutsche Forschungsgemeinschaft (DFG, German Research Foundation) – Project-ID ”458578717”, and by the Federal Ministry of Research, Technology and Space of Germany (BMFTR) for supporting Edge-Cloud AI for DIstributed Sensing and COmputing (AI-DISCO) project (Project-ID ”16ME1127”).
\bibliographystyle{IEEEtran}
\bibliography{ref}

@IEEEtranBSTCTL{IEEEexample:BSTcontrol,
CTLuse_URL = "yes",
CTLuse_forced_etal = "yes",
CTLmax_names_forced_etal = "1",
CTLnames_show_etal= "1" ,
}

@article{tanaka2019recent40,
  title={Recent advances in physical reservoir computing: A review},
  author={Tanaka, Gouhei and Yamane, Toshiyuki and H{\'e}roux, Jean Benoit and Nakane, Ryosho and Kanazawa, Naoki and Takeda, Seiji and Numata, Hidetoshi and Nakano, Daiju and Hirose, Akira},
  journal={Neural Networks},
  volume={115},
  pages={100--123},
  year={2019},
  publisher={Elsevier}
}

@article{zhang2018spiking47,
  title={Spiking echo state convolutional neural network for robust time series classification},
  author={Zhang, Anguo and Zhu, Wei and Li, Juanyu},
  journal={IEEE Access},
  volume={7},
  pages={4927--4935},
  year={2018},
  publisher={IEEE}
}

@article{bloodsample,
  title={Reservoir computing approaches for representation and classification of multivariate time series},
  author={Bianchi, Filippo Maria and Scardapane, Simone and L{\o}kse, Sigurd and Jenssen, Robert},
  journal={IEEE transactions on neural networks and learning systems},
  volume={32},
  number={5},
  pages={2169--2179},
  year={2020},
  publisher={IEEE}
}

@misc{Stream,
      title={Streamlined Deployment for Quantized Neural Networks}, 
      author={Yaman Umuroglu and Magnus Jahre},
      year={2018},
      eprint={1709.04060},
      archivePrefix={arXiv},
      primaryClass={cs.CV},
      howpublished={https://arxiv.org/abs/1709.04060}, 
}

@INPROCEEDINGS{logicnets,
  author={Umuroglu, Yaman and Akhauri, Yash and Fraser, Nicholas James and Blott, Michaela},
  booktitle={2020 30th International Conference on Field-Programmable Logic and Applications (FPL)}, 
  title={LogicNets: Co-Designed Neural Networks and Circuits for Extreme-Throughput Applications}, 
  year={2020},
  volume={},
  number={},
  pages={291-297},
  doi={10.1109/FPL50879.2020.00055}}

@inproceedings{trouvain2020reservoirpy,
  title={Reservoirpy: an efficient and user-friendly library to design echo state networks},
  author={Trouvain, Nathan and Pedrelli, Luca and Dinh, Thanh Trung and Hinaut, Xavier},
  booktitle={International Conference on Artificial Neural Networks},
  pages={494--505},
  year={2020},
  organization={Springer}
}

@article{MI,
  title={Optimizing the echo state network based on mutual information for modeling fed-batch bioprocesses},
  author={Wang, Heshan and Ni, Chunjuan and Yan, Xuefeng},
  journal={Neurocomputing},
  volume={225},
  pages={111--118},
  year={2017},
  publisher={Elsevier}
}

@article{liu2022broad,
  title={Broad echo state network with reservoir pruning for nonstationary time series prediction},
  author={Liu, Wenjie and Bai, Yuting and Jin, Xuebo and Wang, Xiaoyi and Su, Tingli and Kong, Jianlei},
  journal={Computational Intelligence and Neuroscience},
  volume={2022},
  number={1},
  pages={3672905},
  year={2022},
  publisher={Wiley Online Library}
}

@inproceedings{huang2023pruning,
  title={A Pruning Method for Echo State Network Based on Neuron Importance and Iterative Fine-Tuning},
  author={Huang, Xinhui and Zhou, Jian and Yan, Xiaoyong and Ye, Weidu and He, Xin},
  booktitle={2023 China Automation Congress (CAC)},
  pages={1034--1039},
  year={2023},
  organization={IEEE}
}

@article{li2018structure,
  title={Structure optimization for echo state network based on contribution},
  author={Li, Dingyuan and Liu, Fu and Qiao, Junfei and Li, Rong},
  journal={Tsinghua Science and Technology},
  volume={24},
  number={1},
  pages={97--105},
  year={2018},
  publisher={TUP}
}

@article{huang2023semi,
  title={Semi-supervised echo state network with partial correlation pruning for time-series variables prediction in industrial processes},
  author={Huang, Jian and Wang, Fan and Yang, Xu and Li, Qing},
  journal={Measurement Science and Technology},
  volume={34},
  number={9},
  pages={095106},
  year={2023},
  publisher={IOP Publishing}}

@article{henon,
  title={Dynamic memristor-based reservoir computing for high-efficiency temporal signal processing},
  author={Zhong, Yanan and Tang, Jianshi and Li, Xinyi and Gao, Bin and Qian, He and Wu, Huaqiang},
  journal={Nature communications},
  volume={12},
  number={1},
  pages={408},
  year={2021},
  publisher={Nature Publishing Group UK London}
}

@article{othermethod,
  title={Efficient statistical parameter selection for nonlinear modeling of process/performance variation},
  author={Mohammadi, Hassan Ghasemzadeh and Gaillardon, Pierre-Emmanuel and De Micheli, Giovanni},
  journal={IEEE Transactions on Computer-Aided Design of Integrated Circuits and Systems},
  volume={35},
  number={12},
  pages={1995--2007},
  year={2016},
  publisher={IEEE}
}

@article{equation,
  title={Echo state network structure optimization algorithm based on correlation analysis},
  author={Wang, Bowen and Lun, Shuxian and Li, Ming and Lu, Xiaodong},
  journal={Applied Soft Computing},
  volume={152},
  pages={111214},
  year={2024},
  publisher={Elsevier}
}

@article{linear,
  title={Echo state network with multiple delayed outputs for multiple delayed time series prediction},
  author={Yao, Xianshuang and Shao, Yanning and Fan, Siyuan and Cao, Shengxian},
  journal={Journal of the Franklin Institute},
  volume={359},
  number={18},
  pages={11089--11107},
  year={2022},
  publisher={Elsevier}
}

@article{souahlia2020echo_equation,
  title={Echo state network-based feature extraction for efficient color image segmentation},
  author={Souahlia, Abdelkerim and Belatreche, Ammar and Benyettou, Abdelkader and Ahmed-Foitih, Zoubir and Benkhelifa, Elhadj and Curran, Kevin},
  journal={Concurrency and Computation: Practice and Experience},
  volume={32},
  number={21},
  pages={e5719},
  year={2020},
  publisher={Wiley Online Library}
}

@inproceedings{jafari2025ultra,
  title={Ultra-Low Latency and Extreme-Throughput Echo State Neural Networks on FPGA},
  author={Jafari, Atousa and Platzner, Marco},
  booktitle={International Symposium on Applied Reconfigurable Computing},
  pages={179--195},
  year={2025},
  organization={Springer}
}

@article{spearman,
  title={Rethinking the pruning criteria for convolutional neural network},
  author={Huang, Zhongzhan and Shao, Wenqi and Wang, Xinjiang and Lin, Liang and Luo, Ping},
  journal={Advances in Neural Information Processing Systems},
  volume={34},
  pages={16305--16318},
  year={2021}
}

@inproceedings{rakin2019bit,
  title={Bit-flip attack: Crushing neural network with progressive bit search},
  author={Rakin, Adnan Siraj and He, Zhezhi and Fan, Deliang},
  booktitle={Proceedings of the IEEE/CVF International Conference on Computer Vision},
  pages={1211--1220},
  year={2019}
}

@inproceedings{directlogic,
  title={A Two-Stage Approximation Methodology for Efficient DNN Hardware Implementation},
  author={Hadipour, Amir Hossein and Jafari, Atousa and Awais, Muhammad and Platzner, Marco},
  booktitle={2025 IEEE 28th International Symposium on Design and Diagnostics of Electronic Circuits and Systems (DDECS)},
  pages={119--122},
  year={2025},
  organization={IEEE}
}

@article{jafari2025crc,
  title={CRC: Compressed Reservoir Computing on FPGA via Joint HSIC LASSO-based Pruning and Quantization},
  author={Jafari, Atousa and Mohammadi, Hassan Ghasemzadeh and Platzner, Marco},
  journal={WiPiEC Journal-Works in Progress in Embedded Computing Journal},
  volume={11},
  number={1},
  pages={4--4},
  year={2025}
}

@inproceedings{prune,
  title={Mix-and-Match Pruning: Globally Guided Layer-Wise Sparsification of DNNs},
  author={Monachan, Danial and Nazari, Samira and Taheri, Mahdi and Azarpeyvand, Ali and Krstic, Milos and H{\"u}bner, Michael and Herglotz, Christian},
  booktitle={2026 12th International Conference on Computing and Artificial Intelligence (ICCAI)},
  year={2026},
  organization={IEEE},
  note={In press},
  address={Okinawa, Japan}
}

@inproceedings{hawx,
  title={HAWX: A Hardware-Aware Framework for Fast and Scalable Approximation of DNNs},
  author={Nazari, Samira and Almasi, Mohammad Saeed and Taheri, Mahdi and Azarpeyvand, Ali and Mokhtari, Ali and Mahani, Ali and Herglotz, Christian},
  booktitle={2026 Design, Automation and Test in Europe Conference (DATE)},
  year={2026},
  organization={IEEE},
  note={In press}
}
\end{document}